\documentclass[%
reprint, 
superscriptaddress,
 amsmath,amssymb,
 aps, physrev,
pra,
]{revtex4-2}
\usepackage{graphicx}
\usepackage{layouts}
\usepackage{braket} 
\usepackage{xcolor} 

\begin{document}
\title{Probing the Quantum Capacitance of Rydberg Transitions of Surface Electrons on Liquid Helium via Microwave Frequency Modulation}

\author{Asher Jennings}
    \email[]{asher.jennings@riken.jp}
\affiliation{RIKEN Center for Quantum Computing, 2-1 Hirosawa, Wako, Saitama, 351-0198, Japan}
\author{Ivan Grytsenko}
\affiliation{RIKEN Center for Quantum Computing, 2-1 Hirosawa, Wako, Saitama, 351-0198, Japan}
\author{Yiran Tian}
\affiliation{RIKEN Center for Quantum Computing, 2-1 Hirosawa, Wako, Saitama, 351-0198, Japan}
\affiliation{Institute of Physics, Kazan Federal University, 16a Kremlyovskaya St. Kazan, 420008, Republic of Tatarstan, Russian Federation}
\author{Oleksiy Rybalko}
\affiliation{RIKEN Center for Quantum Computing, 2-1 Hirosawa, Wako, Saitama, 351-0198, Japan}
\affiliation{B. Verkin Institute for Low Temperature Physics and Engineering of the National Academy of Sciences of Ukraine, 47 Nauky Ave. Kharkiv, 61103, Ukraine}
\author{Jun Wang}
\affiliation{RIKEN Center for Quantum Computing, 2-1 Hirosawa, Wako, Saitama, 351-0198, Japan}
\author{Itay Josef Barabash}
\affiliation{RIKEN Center for Quantum Computing, 2-1 Hirosawa, Wako, Saitama, 351-0198, Japan}
\affiliation{McGill University, School of Computer Sciencem 805 Sherbrooke Street West, Montreal, Quebec H3A 2K6, Canada}
\author{Erika Kawakami}
    \email[]{e2006k@gmail.com}
\affiliation{RIKEN Center for Quantum Computing, 2-1 Hirosawa, Wako, Saitama, 351-0198, Japan}
\affiliation{RIKEN Cluster for Pioneering Research, 2-1 Hirosawa, Wako, Saitama, 351-0198, Japan.}

\date{\today} 

\begin{abstract}
We present a method for probing the quantum capacitance associated with the Rydberg transition of surface electrons on liquid helium using radio-frequency (RF) reflectometry. Resonant microwave excitation of the Rydberg transition induces a redistribution of image charges on capacitively coupled electrodes, giving rise to a quantum capacitance originating from adiabatic state transitions and the finite curvature of the energy bands.
By applying frequency-modulated resonant microwaves to drive the Rydberg transition, we systematically measured a capacitance sensitivity of 0.34~aF/$\sqrt{\mathrm{Hz}}$ for our RF reflectometry scheme. This sensitivity is sufficient to detect the Rydberg transition of a single electron, offering a scalable pathway toward qubit readout schemes based on surface electrons on helium.
\end{abstract}

\keywords{electron on helium, cryogenic LC circuit, Rydberg state, quantum state read out}

\maketitle
\section{Introduction}
Surface electrons (SEs) floating on liquid helium form an exceptionally pure two-dimensional electron system, providing a promising platform for qubit implementation. Several theoretical proposals have been based on this system~\cite{Lyon2006,Platzman1999,Lea2000,Schuster2010,Dykman2023-hx,Zhang2012,Jennings2024-sb}. The orbital states of the SEs perpendicular to the liquid helium surface are typically referred to as Rydberg states~\cite{Monarkha2004-un, Andrei1997Two-DimensionalSubstrates}, with the energy levels given by  \( E_{n_z} =  -  R_\infty \left(\frac{\Lambda}{4 }\right)^2  \frac{1}{ n_z^2} \), where \( R_{\infty}\) is the Rydberg constant, \( n_z \) is the quantum number of the Rydberg state, and \( \Lambda=0.0272 \) for liquid helium-4.

The initial proposal for quantum computing using SEs defined qubits using the two lowest Rydberg states~\cite{Platzman1999,Dykman2003}. Later proposals considered the use of electron spin as qubit states more advantageous, due to its predicted coherence time of several seconds~\cite{Schuster2010,Lyon2006}. However, direct spin readout is challenging because of its small magnetic moment. To address this, indirect detection schemes have been proposed,  coupling the spin state to either the orbital state associated with motion parallel to the helium surface (in-plane orbital state)~\cite{Schuster2010} or the Rydberg state~\cite{Kawakami2023-vf}, enabling spin detection through these coupled degrees of freedom.

For the former approach, experimental efforts have successfully demonstrated the coupling of the in-plane orbital state of a single electron to a microwave superconducting resonator~\cite{Koolstra2019-mq}. Among related platforms, solid neon has shown more rapid progress, with recent experiments demonstrating high-fidelity orbital qubits~\cite{Zhou2022-nk,Zhou2023-iw,Li2025-em}. In contrast, while qubits have not yet been realized on liquid helium, the system offers unique scalability advantages, particularly due to the absence of surface roughness—a key limitation in solid-state substrates.

We focus on the latter approach, which couples the spin state to the Rydberg state. In this scheme, the occurrence of a Rydberg transition reflects the spin state and enables spin readout~\cite{Kawakami2023-vf}. This readout scheme is analogous to spin-to-charge conversion in quantum dot-based systems, such as Pauli spin blockade in double quantum dots (DQDs)~\cite{Ono2002-as}, where the spin state determines whether a charge transition occurs. In this analogy, the ground and first excited Rydberg states of our system are mapped onto the DQD charge states. Instead of microwave superconducting resonators, radio-frequency (RF) LC tank circuits are used to detect the Rydberg transition, inspired by techniques in quantum dot systems where charge transitions in DQDs are detected via RF reflectometry~\cite{Aassime2001-gg,Brenning2006-wc,Ahmed2018-he,Gonzalez-Zalba2015,Oakes2023-ys,Ibberson2021-uq,Apostolidis2024-kr}.  While this method does not allow coherent coupling to microwave photons, it offers better scalability due to the smaller footprint of LC circuits. As a proof-of-concept, we demonstrate the detection of Rydberg transitions from many electrons using an LC tank circuit, leveraging its high sensitivity as a step toward single-electron detection.

\section{Experimental setup}

\begin{figure}[h]
\centering
\includegraphics[width=1\linewidth]{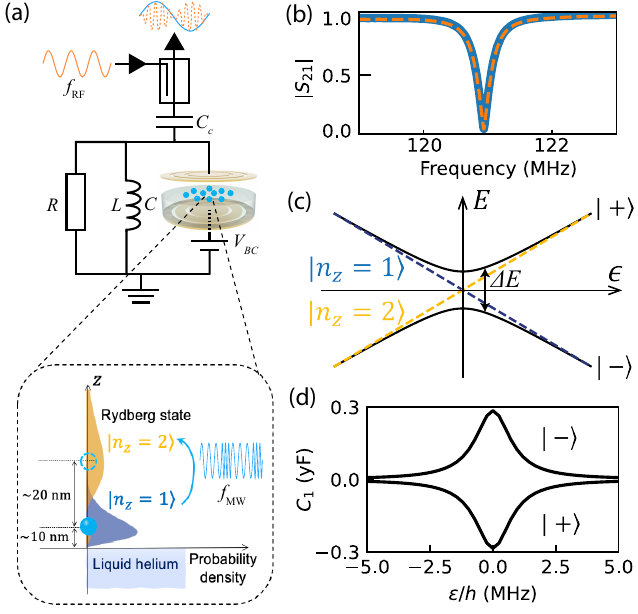}
\caption{(a) Schematic of the LC circuit incorporating the Corbino electrodes as a capacitor, with $10^7$ SEs (light blue circles) floating on liquid helium-4 (semi-transparent blue) half-way between. The capacitor is integrated into an LC circuit housed in a leak-tight cell on the mixing chamber plate of  a dilution refrigerator. See the main text for details on the components of the LC circuit. The inset shows the probability density of an electron in the Rydberg-ground state (blue) and in the Rydberg-1st-excited state (yellow), showing its average position from the liquid helium surface. The light blue curves show the FM-MW signal used to induce the Rydberg transition, with carrier frequency \( f_\mathrm{MW} \) and modulation frequency \( f_\mathrm{mf} \) as defined in Eq.~\ref{eq:FM-MW}. Returning to the main panel, the rectangle represents an RF coupler. The orange curves represent the incident RF signal with frequency \( f_\mathrm{RF} \), while the dashed orange curves depict the reflected RF signal. A light blue sinusoidal curve overlapping the orange dashed curves indicates that the reflected signal is amplitude-modulated with \( f_\mathrm{mf} \). (b) Normalized magnitude of the reflectance of the LC circuit measured with a VNA (blue dots) at 150~mK, along with a fit (orange line) without electrons and without MW irradiation. See the main text for details of the fitting results. (c) Schematic energy diagram as a function of the detuning from the Rydberg resonance \(\epsilon\), showing the energy levels of a single electron under microwave irradiation (solid black lines) and without irradiation (dashed lines). (d) Quantum capacitance of a single electron $C_1^{\pm}$ as a function of \( \epsilon/h \) for the eigenstates \( \ket{\pm} \), with the Rydberg transition rate \( 2t_c/h = 0.83~\mathrm{MHz} \).
}\label{fig1}
\end{figure}

We built a parallel LC circuit~\cite{Ahmed2018-he,Gonzalez-Zalba2015,Oakes2023-ys,Ibberson2021-uq,Apostolidis2024-kr}, as shown in Fig.~\ref{fig1}(a). In our experimental setup, two sets of parallel plate electrodes with a Corbino geometry~\cite{Iye1980-zn,Mehrotra1987,Wilen1988-ui} are used, with the plates separated by $D=2$~mm. Both the top and bottom plates consist of three concentric electrodes, each with an area of approximately $0.5$~cm$^2$, forming a capacitor. The top central electrode is connected to a niobium spiral inductor microfabricated on a sapphire substrate, which is used to suppress resonator losses thanks to its low dielectric loss and superconducting properties~\cite{Colless2013-fc}. In addition, low-pass filters are implemented on the DC voltage lines to minimize noise coupling to the resonator (Appendix~\ref{sec:details_circuit}). The LC circuit resonance without electrons, measured using a vector network analyzer (VNA), is shown in Fig.~\ref{fig1}(b). From separate measurements, the inductance was determined to be \( L = 708~\mathrm{nH} \). Fitting Eq.~\ref{eq:Gamma_quality_factor}~\cite{Probst2015-gs,probst2024resonator} gives a loaded quality factor \( Q_\mathrm{tot} = 311 \), an external quality factor \( Q_\mathrm{ext} = 648 \), a coupling capacitance \( C_\mathrm{c} = 0.315~\mathrm{pF} \), a total capacitance \( C = 2.131~\mathrm{pF} \), and an effective resistance \( R = 321~\mathrm{k}\Omega \). The resonance frequency is \( f_0 =\frac{1}{2\pi \sqrt{LC_\mathrm{t}}}= 120.946~\mathrm{MHz} \), where $C_\mathrm{t}=C+C_\mathrm{c}$. $R$ accounts for distributed circuit losses~\cite{Ahmed2018-he,Oakes2023-ys,Ibberson2021-uq} and \( C \) includes contributions from both parasitic and Corbino capacitance after the coupling capacitor.

Microwaves (MWs) are applied to the SEs via a TEM waveguide extending from room temperature into the leak-tight cell, positioned between the top and bottom electrodes. The Rydberg transition energy, \( h f_\mathrm{Ry} \) where \( h \) is Planck's constant, can be tuned by the electric field perpendicular to the helium surface via the Stark effect~\cite{Grimes1976SpectroscopyHelium}. The MWs resonantly drive transitions between the Rydberg-ground state and the first excited state~\cite{Grimes1974-fg,Lambert1979ElectronsHelium,Collin2002MicrowaveHelium,Konstantinov2007-tw,Kawakami2019} when the MW frequency \( f_\mathrm{MW} \) matches the Rydberg transition frequency \( f_\mathrm{Ry} \). The bottom center electrode is labeled \(\mathrm{BC}\), and the combination of the bottom middle and outer electrodes is referred to as \(\mathrm{BG}\). DC voltages of \( V_\mathrm{BC} = 12~\mathrm{V} \) and \( V_\mathrm{BG} = -90~\mathrm{V} \) are applied, while all other electrodes are grounded. Note that this charge configuration is chosen to prevent interference from lateral modes of motion, known as plasmons~\cite{Grimes1976-kr}, with the Rydberg measurement. These settings produce a nominal perpendicular electric field \( E_z = V_\mathrm{BC}/D \) to the helium surface. This field sets the peak Rydberg transition frequency to approximately \( f_\mathrm{Ry}^0 \approx 165~\mathrm{GHz} \).
When an electron is excited from the Rydberg ground state to the Rydberg first-excited state, their average distance from the liquid surface increases by \( d = 20~\mathrm{nm} \) (Fig.~\ref{fig1}(a)), altering the image charge induced on the top plate electrode by  \( \Delta q = \frac{d}{D}e = 10^{-5} e \), where \( e \) is the elementary charge~\cite{Kawakami2019}. Previously, such changes induced by many electrons were detected as a current~\cite{Kawakami2019} and as a voltage~\cite{Kawakami2021}. In this work, we detect such changes as a quantum capacitance~\cite{Smith1985-zz,Luryi1988-ek,Duty2005-rq,Mizuta2017-xl,Gonzalez-Zalba2015,Crippa2019-ir,Vigneau2023-yi} with an LC circuit using RF reflectometry. This capacitance arises from the finite curvature of the Rydberg energy bands when the Rydberg transition is resonantly driven by microwaves. We assess the circuit’s performance by measuring its capacitance sensitivity in a controlled manner using a frequency-modulated microwave (FM-MW) technique.

\section{Quantum capacitance}

Here, we focus on the two lowest Rydberg states, and the Hamiltonian for an electron can be written as~\cite{Kawakami2023-vf}
\begin{equation}
    H = \frac{h f_{\mathrm{Ry}}}{2} \sigma_z + t_c \, \sigma_x \cos\left(2\pi f_{\mathrm{MW}} t\right),
\end{equation}
where $\sigma_z = \ket{n_z=2}\bra{n_z=2} - \ket{n_z=1}\bra{n_z=1}$ and 
$\sigma_x = \ket{x+}\bra{x+} - \ket{x-}\bra{x-}$. 
Here, $\ket{n_z=2}$ denotes the Rydberg first-excited state, 
$\ket{n_z=1}$ the Rydberg ground state, and 
$\ket{x\pm} = \frac{1}{\sqrt{2}} \left( \ket{n_z=2} \pm \ket{n_z=1} \right)$. The second term represents MW irradiation at frequency $f_{\mathrm{MW}}$, which induces transitions between $\ket{n_z=2}$ and $\ket{n_z=1}$. 
By moving to the rotating frame at frequency $f_{\mathrm{MW}}$, the Hamiltonian is transformed to
\begin{equation}
    H_\mathrm{r} = \frac{\epsilon}{2} \sigma_z + t_c \, \sigma_x,
\end{equation}
where $\epsilon = h (f_{\mathrm{Ry}} - f_{\mathrm{MW}})$ is the detuning from the Rydberg resonance, and $2t_c/h$ is the Rydberg transition rate. 
The eigenstates of $H_\mathrm{r}$ are denoted by $\ket{\pm}$ and their energies are given by $E_\pm= \pm \frac{1}{2}\sqrt{\epsilon^2+(2t_c)^2}$ . 
Figure~\ref{fig1}(c) shows a schematic energy diagram of this two-level system. For RF reflectometry, an incident signal $ V_\mathrm{RF} \cos(2\pi f_\mathrm{RF} t)$ is injected into the LC circuit, and the reflected signal is measured using a spectrum analyzer. Throughout this work, we use a frequency of $f_\mathrm{RF} = 120.94~\mathrm{MHz}$, which is set to the LC resonance frequency $f_0$ in the presence of SEs.



\begin{figure}
\centering
\includegraphics{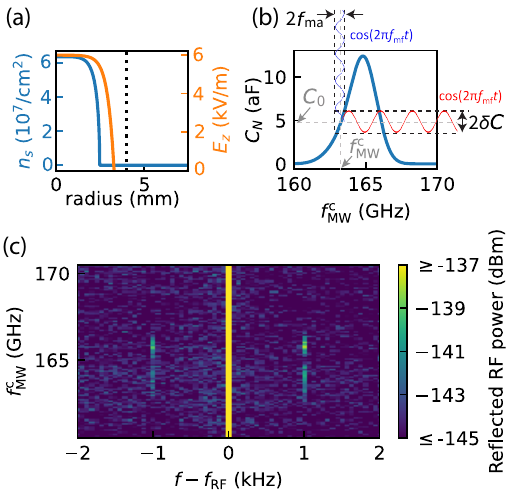}
\caption{
(a) Calculated saturated electron density $n_s$ (blue, left axis) and vertical electric field $E_z$ (orange, right axis) as a function of radial distance from the center of the Corbino electrodes, for $V_\mathrm{BC} = 12$~V and $V_\mathrm{BG} = -90$~V. Given that the radius of the central electrode is 4~mm (vertical dotted line), all electrons are confined well within this region. Only the positive values of the electric field are shown for clarity.
(b) Quantum capacitance of many electrons $C_N$ as a function of the microwave frequency $f_\mathrm{MW}$ with the Rydberg transition rate $2t_c/h=0.83$~MHz. The FM-MW signal in Eq.~\ref{eq:FM-MW} induces a modulation in the quantum capacitance at \( f_{\mathrm{mf}} \), as described in Eq.~\ref{eq:C_N}. (c) Reflected RF power measured with a spectrum analyzer as a function of the MW carrier frequency $f_\mathrm{MW}^c$. The FM modulation parameters are \( f_\mathrm{mf} = 1~\mathrm{kHz} \) and \( f_\mathrm{ma} = 768~\mathrm{MHz} \). In all experiments presented in the main text, the microwave power irradiated on the electrons is fixed (see Appendix~\ref{sec:power_dependence} for the estimated value).
Sideband signals appear at $f = f_\mathrm{RF} \pm f_\mathrm{mf}$ around $f_\mathrm{MW}^c = 165$~GHz, associated with the Rydberg transition. Only every 250\textsuperscript{th} data point is plotted along $f$ for clarity. The sharp signals at $f-f_\mathrm{RF}=\pm 1$~kHz lie on retained points, with no significant features missed due to downsampling.}
\label{fig2}
\end{figure}

When an electron is in the state \( \ket{\pm} \), the induced image charge on the top plate electrode is given by
\( Q_1^{\pm} = \Delta q \, |\braket{n_z=2|\pm}|^2 = \frac{\Delta q}{2} \left( 1 \mp \frac{\epsilon}{\Delta E} \right) \),
where \( \Delta E = E_+ - E_- \). The corresponding quantum capacitance (Appendix~\ref{sec:quantum_C}) is proportional to the curvature of the energy bands:
\begin{equation}
   C_1^{\pm}(\epsilon) = \Delta q \, \frac{dQ_1^{\pm}}{d\epsilon} = \mp \Delta q^2 \frac{(2t_c)^2}{2\Delta E^3} = \mp \Delta q^2 \frac{d^2 E_\pm}{d \epsilon^2},
\end{equation}
and is shown in Fig.~\ref{fig1}(d). Introducing the population difference between the \( \ket{+} \) and \( \ket{-} \) states as \( \chi \), the quantum capacitance of a single electron is expressed as
\begin{equation}
   C_1(\epsilon) = \chi \, \Delta q^2 \frac{(2t_c)^2}{2\Delta E^3},\label{eq:C_1}
\end{equation}
where $\chi\approx t_c/k_BT =1.2 \times 10^{-4}$  for $2t_c/h=0.83$~MHz and $T=160$~mK is the temperature (see Sec.~\ref{sec:LZ} for the derivation of $2t_c$).   To capture the quantum capacitance arising from a large number of SEs, we convolute the single-electron quantum capacitance \( C_1(\epsilon) \) with the distribution \( n(\epsilon) \), which represents the number of electrons per energy detuning (Appendix~\ref{sec:Ez_fRy_distribution}):
\begin{equation}
    C_N(\epsilon^0) =  \int_{-\infty}^{+\infty} C_1(\epsilon^0 -\epsilon)\, n(\epsilon)\, d\epsilon , \label{eq:C_N}
\end{equation}

\noindent
where the detuning for the ensemble of SEs is defined as \( \epsilon^0 = h  (f^0_\mathrm{Ry} - f_\mathrm{MW}) \). Numerically calculated $C_N(f_\mathrm{MW})$ is plotted in Fig.~\ref{fig2}(b). As shown in Fig.~\ref{fig2}(a), both the electron density \( n_s \) and the perpendicular electric field \( E_z \) experienced by the electrons vary with radial distance from the center, leading to a corresponding variation in \( f_\mathrm{Ry} \). This spatial inhomogeneity results in an asymmetry in \( C_N(f_\mathrm{MW}) \) (Appendix~\ref{sec:Ez_fRy_distribution}).

\section{Frequency modulation}
To probe the system, we irradiate the SEs with a FM-MW signal. The time-dependent frequency of the FM-MW is given by

\begin{equation}
f_\mathrm{MW}(t) = f_\mathrm{MW}^\mathrm{c} + f_\mathrm{ma} \cos(2\pi f_\mathrm{mf} t),
\label{eq:FM-MW}
\end{equation}

\noindent
where \( f_\mathrm{MW}^\mathrm{c} \) is the carrier frequency, \( f_\mathrm{ma} \) is the modulation amplitude, and \( f_\mathrm{mf} \) is the modulation frequency. Under FM modulation, the detuning \( \epsilon^0 \) is modulated at frequency \( f_\mathrm{mf} \) with amplitude \( h f_\mathrm{ma} \), inducing a time-dependent variation in the quantum capacitance \( C_N \). On either side of the peak of \( C_N \), and for sufficiently small \( f_\mathrm{ma} \), \( C_N \) is approximately linear in \( \epsilon^0 \) and can be expressed as
\begin{equation}
    C_N(t) \approx C_0 + \delta C \cos(2\pi f_\mathrm{mf} t),
\end{equation}
where \( C_0 \) is the average capacitance at a fixed detuning, and \( \delta C \) is the amplitude of capacitance modulation (Fig.~\ref{fig2}(b), Appendix~\ref{sec:quantum_C}). Expressing the reflection coefficient \( \Gamma_\mathrm{ref} \) as a function of capacitance (Appendix~\ref{sec:reflectance_change}), the time-dependent reflection coefficient under FM-MW irradiation becomes
\begin{equation}
        \Gamma_\mathrm{ref}(C_\mathrm{t}+C_N(t))  
         \approx  \Gamma_\mathrm{ref}(C_\mathrm{t}) 
- j\frac{2Q_\mathrm{tot}^2}{Q_\mathrm{ext}} \frac{C_0 + \delta C \cos(2\pi f_\mathrm{mf} t)}{C_\mathrm{t}}.
\label{eq:GammaDiff}
\end{equation}

\noindent
Accordingly, the reflected signal is given by $ V_\mathrm{o}=\Gamma_\mathrm{ref}(C_\mathrm{t} + C_N(t))\, G\, V_\mathrm{RF} \cos (2\pi f_\mathrm{RF} t)$, where \( V_\mathrm{RF} \) is the voltage amplitude of the RF signal at the top-plate electrode, and \( G = 41 \) is the total gain from the output of the LC resonator to the input of the spectrum analyzer. A portion of the reflected signal is amplitude modulated at frequency \( f_\mathrm{mf} \), and thus contains sideband components at \( f_\mathrm{RF} \pm f_\mathrm{mf} \). The amplitude of the sideband signals is given by
\begin{equation}
V_\mathrm{s}  = G  \frac{Q_\mathrm{tot}^2}{Q_\mathrm{ext}} \frac{  |\delta C|}{C_\mathrm{t}} V_\mathrm{RF}.
\label{eq:Vs}
\end{equation}

\noindent Note that the Rydberg transition can also be detected under continuous-wave microwave irradiation without FM~\cite{jennings_inprep}, but this requires phase-sensitive detection (Appendix~\ref{sec:quantum_C}). In contrast, FM shifts the quantum capacitance signal to distinct sideband frequencies, enabling detection via amplitude measurements. Furthermore, varying FM parameters allows systematic control over the signal strength, enabling more controlled and quantitative characterization.

Figure~\ref{fig2}(c) shows the reflected RF power measured with the spectrum analyzer with \( f_\mathrm{mf} = 1~\mathrm{kHz} \). Sideband peaks are observed at \( f=f_\mathrm{RF} \pm 1~\mathrm{kHz} \) and around \( f_\mathrm{MW}^\mathrm{c} = 165~\mathrm{GHz} \). At exact Rydberg-resonance, \( f_\mathrm{MW}^\mathrm{c} = 165~\mathrm{GHz} \), the signal vanishes because \( \delta C= 0 \) at this point~\footnote{Sideband signals at \( f_\mathrm{RF} \pm 2~\mathrm{kHz} \) were expected to be observed at this point, but they were extremely small and difficult to distinguish from the noise.}. The signal exhibits peaks at \( \pm 1~\mathrm{GHz} \) offsets from the resonance, as shown in Fig.~\ref{fig2}(c) and Fig.~\ref{fig3}(a). The peak at \( 166~\mathrm{GHz} \) is larger than that at \( 164~\mathrm{GHz} \), which is attributed to the steeper slope in $C_N$ on the high-frequency side.

\section{Capacitance sensitivity}

\begin{figure}
\centering
\includegraphics[]{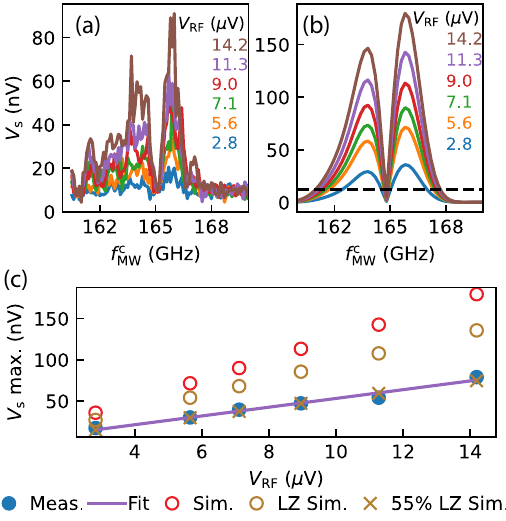}
\caption{(a) Sideband amplitude $V_\mathrm{s}$ at $f = f_\mathrm{RF} + f_\mathrm{mf}$ as a function of the MW carrier frequency $f_\mathrm{MW}$ for different $V_\mathrm{RF}$, measured with $f_\mathrm{mf} = 1$~kHz, $f_\mathrm{ma} = 768$~MHz.  (b) Simulated counterpart of (a). At low $V_\mathrm{RF}$, the signal drops below the noise level $V_\mathrm{n}=12$~nV (dashed black line). (c) Blue points (red circles) represent the maximum sideband amplitude extracted from the experimental data in (a) (simulation in (b)) as a function of \( V_\mathrm{RF} \). Bronze circles (crosses) represent the simulation results including Landau--Zener (LZ) effects with the saturated electron density (55\% of the saturated electron density) (see Sec.~\ref{sec:LZ}). To determine the experimental maxima, a Gaussian rolling average with a width of 5 points was applied. The purple line is a fit of Eq.~(\ref{eq:Vs}) to the experimental data.}\label{fig3}
\end{figure}

Figures~\ref{fig3}(a) and (b) show the measured and simulated sideband amplitude $V_\mathrm{s}$ at $f=f_\mathrm{RF}+f_\mathrm{mf}$ as a function of $f_\mathrm{MW}^c$ for different \( V_\mathrm{RF} \), respectively. This simulation is performed by calculating the time-domain modulated $C_N(t)$ using Eq.~\ref{eq:C_N} and Eq.~\ref{eq:FM-MW} and taking the Fourier transform of the reflected signal $V_\mathrm{o} $ (Appendix ~\ref{sec:Ez_fRy_distribution}). The simulation is performed without using any parameters extracted from the experimental data, except for the Rydberg transition rate, $2t_c/h = 0.83$~MHz, which is taken from Fig.~\ref{fig4}(b). Figure~\ref{fig3}(c) shows the maximum sideband amplitude measured and simulated around \( f_\mathrm{MW}^c = 166~\mathrm{GHz} \) as a function of \( V_\mathrm{RF} \). The experimental amplitude is smaller than the simulated one even if the effect of the Landau--Zener transition discussed in Sec.~\ref{sec:LZ} is also included (bronze circles in Fig.~\ref{fig3}(c)), which could be attributed to a reduction in the electron density below the saturated value due to electron loss, or a higher electron temperature compared to the thermometer, which is located outside the experimental cell and reads approximately $T=$160~mK in the data shown in this manuscript. The bronze crosses in Fig.~\ref{fig3}(c) indicate that the experimental value agrees with the simulation if the electron density is assumed to be 55\% of the saturated value.

The purple line in Fig.~\ref{fig3}(c) shows fitting Eq.~\ref{eq:Vs} to the experimental data yields \(
\frac{|\delta C|}{C_\mathrm{t}} = 8.6 \times 10^{-7}
\). From this, we obtain  \( |\delta C| = 2.1~\mathrm{aF} \). The capacitance sensitivity is given by
\begin{equation}
    S_c = \frac{|\delta C| \, V_\mathrm{n}}{ \sqrt{B} \, V_\mathrm{s} } 
    = \frac{Q_\mathrm{ext} C_\mathrm{t} V_\mathrm{n}}{G Q_\mathrm{tot}^2 \sqrt{B} V_\mathrm{RF}} 
    = 0.34 ~\mathrm{aF/\sqrt{Hz}}
\end{equation}

\noindent for $V_\mathrm{RF} = 14~\mathrm{\mu V}$, where the voltage noise is $V_\mathrm{n} = 12$~nV and the measurement bandwidth is $B = 1$~Hz. These results demonstrate that RF reflectometry is well suited for detecting the Rydberg transition of a single SE, as the resulting capacitance change in nanoscale devices is expected to be 60~aF~\cite{Kawakami2023-vf}. With a measurement bandwidth of 10~Hz, detection with a signal-to-noise ratio of approximately 1 is feasible. Sensitivity could be further enhanced by using lossless variable capacitors based on ferroelectric materials such as STO~\cite{Apostolidis2024-kr} to achieve critical coupling. At critical coupling, where \( Q_\mathrm{int} = Q_\mathrm{ext} \), the sensitivity coefficient scales as \( S_c \propto 1/Q_\mathrm{int} \). The internal quality factor \( Q_\mathrm{int} \) can be increased either by increasing the resonance frequency \( f_0 \) or by suppressing circuit losses, i.e., increasing \( R \) (Appendix~\ref{sec:appendix_compare}).

\section{Landau--Zener transition \label{sec:LZ}}

\begin{figure}[ht!]
\centering
\includegraphics[width=1\linewidth]{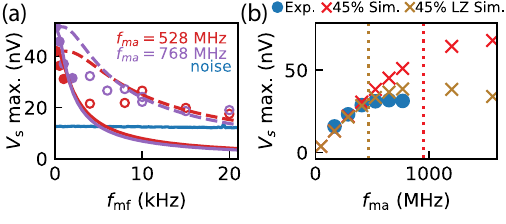}
\caption{ (a) Maximum sideband amplitude as a function of FM frequency \( f_\mathrm{mf} \) for \( f_\mathrm{ma} = 528 \)~MHz (pink circles) and \( 768 \)~MHz (purple circles). Solid lines are a joint fit of both $f_\mathrm{ma}$ only to the data in the domain \( f_\mathrm{mf} \leq 3 \)~kHz (filled circles), while dashed lines are a joint fit for the domain up to \( f_\mathrm{mf} = 20 \)~kHz (solid and open circles). The extracted Rydberg transition rates are \( 2t_c/h = 0.83 \pm 0.29 \)~MHz and \( 1.74 \pm 0.22 \)~MHz, for the two fit domains respectively, with 99\% confidence intervals. The blue line indicates the measured noise level. Data for \( f_\mathrm{mf} > 20 \)~kHz fall below the noise floor and are omitted. (b) Maximum sideband amplitude as a function of FM amplitude \( f_\mathrm{ma} \), with \( f_\mathrm{mf} = 1 \)~kHz. Blue points (red crosses) represent the experimental data (simulation data). Bronze crosses represent the simulation results including Landau--Zener (LZ) effects. The dotted red and bronze lines show the values up to which there is linear relationship for $f_{ma}$ and $\delta C$, with and without the LZ transition, respectively.  The simulation results use 45\% of the saturated electron density. In both (a) and (b), \( V_\mathrm{RF} = 9~\mu\mathrm{V} \).}
\label{fig4}
\end{figure}

For a coupled two-level system with a modulated detuning, Landau--Zener (LZ) transitions need to be considered~\cite{Shevchenko2010,Landau1932,Zener1932,Stuckelberg1932}. In this work, although the detuning is modulated continuously, it is sufficient to consider only a single passage through \(\epsilon = 0\), as \(f_\mathrm{mf}\) is much slower than the typical relaxation rate of the Rydberg state of electrons on liquid helium (1~MHz~\cite{Monarkha2007-el,Monarkha2006-gi,Kawakami2021}), allowing the system to reach thermal equilibrium before the next passage. Starting from the state \( \ket{-} \) at $|\epsilon| \gg 0$, the system undergoes a LZ transition when passing through \( \epsilon = 0 \), switching to \( \ket{+} \) with probability \( P_{\mathrm{LZ}} = \exp(-2\pi\delta) \), where \( \delta = (2t_c)^2 / 4 f_{\mathrm{ma}} f_{\mathrm{mf}} \), or remaining in \( \ket{-} \) with probability \( 1 - P_{\mathrm{LZ}} \).  Only the population that stays in \( \ket{-} \) contributes to the change in quantum capacitance, and thus to the observed signal. An increase in \( f_{\mathrm{mf}} \) enhances the LZ transition probability \( P_{\mathrm{LZ}} \), which in turn suppresses the signal amplitude. Figure~\ref{fig4}(a) shows the experimentally measured sideband peak amplitude as a function of \( f_{\mathrm{mf}} \). The data were fitted using the expression \( a(1 - P_{\mathrm{LZ}}) \), where \( a \) denotes the amplitude in the absence of LZ transitions and \( 2t_c \) was treated as a fitting parameter. From the fit, we extracted a Rydberg transition rate of \( 2t_c /h = 0.83~\mathrm{MHz} \) (\( 1.74~\mathrm{MHz} \)), using data points with \( f_{\mathrm{mf}} \leq3~\mathrm{kHz} \) (\( f_{\mathrm{mf}} \leq 20~\mathrm{kHz} \)). Slightly different Rydberg transition rates depending on \( f_{\mathrm{mf}}\) may arise from variation in the effective microwave power, possibly due to frequency-dependent transmission characteristics in the frequency multiplier or the waveguide. Except for Fig.~\ref{fig4}(a), all experimental data were acquired at \( f_{\mathrm{mf}} = 1~\mathrm{kHz} \), and adopting \( 2t_c/h = 0.83~\mathrm{MHz} \) in the analysis yields better consistency with those data. Therefore, we adopt \( 2t_c/h = 0.83~\mathrm{MHz} \) throughout the main text. Additionally, as the FM modulation amplitude \( f_{\mathrm{ma}} \) increases, the signal begins to be suppressed, as seen in Fig.~\ref{fig4}(b). Although this behavior is partly due to the system leaving the linear region of \( C_N \) at high \( f_\mathrm{ma} \), the earlier onset of the signal drop is consistent with the effect of LZ transitions. Note that the experiments shown in Fig.~\ref{fig4}(b) and Fig.~\ref{fig:power_dependence} were conducted at a different time from those in Fig.~\ref{fig3} and Fig.~\ref{fig4}(a). Even under the same conditions, the signal was smaller, which we attribute to a loss of electrons. In the simulations in  Fig.~\ref{fig4}(b) and Fig.~\ref{fig:power_dependence}, this effect was included by assuming that the number of electrons had decreased to 45\% of the saturated density.

\section{Conclusion}

We have demonstrated the measurement of quantum capacitance induced by the Rydberg transition of surface electrons on liquid helium using FM-MW. The observed signal shape and amplitude are well explained by a model incorporating quantum capacitance and LZ transitions induced by FM. FM enables amplitude-based detection without phase-sensitive techniques and allows systematic tuning of the signal strength for quantitative characterization. Our LC circuit, which incorporates a microfabricated superconducting coil and employs filters to suppress external losses, achieves a high quality factor with minimal dissipation. Consequently, we attained a capacitance sensitivity of \( 0.34~\mathrm{aF}/\sqrt{\mathrm{Hz}} \), sufficient to resolve the Rydberg transition of a single electron and promising for future readout of qubit states.

\section*{Acknowledgements}
This work was supported by the RIKEN Hakubi Program, the RIKEN Center for Quantum Computing, JST FOREST, and the Yazaki Foundation. We thank Prof. Denis Konstantinov and his team for valuable discussions.

\appendix
\section{Details of the circuits\label{sec:details_circuit}}

\begin{figure}[h!]
    \centering
    \includegraphics[width=1\linewidth]{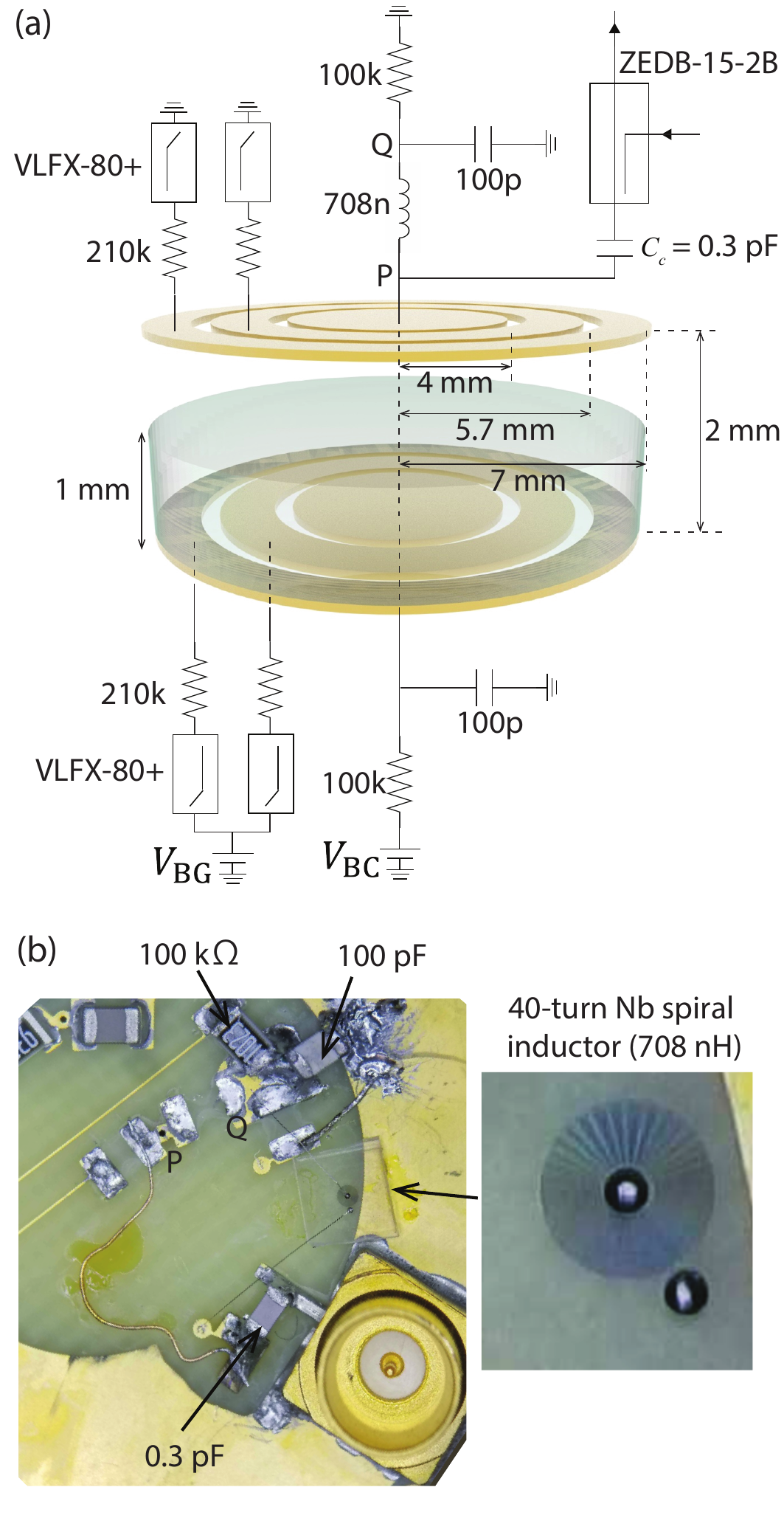}
    \caption{(a) The VLFX-80+ low-pass filters from Mini-Circuits attenuate signals above 80 MHz to reduce high-frequency noise and minimize losses in the LC resonant circuit. All electrical components shown are mounted on the top and bottom plates of the Corbino electrodes inside the experimental cell, except for the ZEDB-15-2B directional coupler, which is located outside the cell at the mixing chamber plate.
(b) Photograph of the top-plate electrode. The inset shows the microfabricated 40-turn Nb spiral inductor on a sapphire substrate. Points $P$ and $Q$ in (a) correspond to the same points as $P$ and $Q$ in (b), respectively.}
    \label{fig:circuit_details}
\end{figure}

The circuit details inside the leak-tight cell, including the RF coupler, are shown in Fig.~\ref{fig:circuit_details}. The incident signal is generated by a signal generator (E8267D, Keysight), passed through cryogenic attenuators totaling \(-20~\mathrm{dB}\) (\(-10~\mathrm{dB}\) at the \(1~\mathrm{K}\) stage and \(-10~\mathrm{dB}\) at the \(100~\mathrm{mK}\) stage), and applied to the coupled port of the RF coupler (Mini-Circuits, ZEDB-15-2B). The resulting signal \( V_\mathrm{RF} \cos(2\pi f_\mathrm{RF} t) \) is delivered to the LC circuit. The reflected signal is returned from the coupler's output port and amplified by a cryogenic amplifier (CITLF-3, Cosmic Microwaves) operating at 4~K and is measured as $V_\mathrm{o}$ with a spectrum analyzer (FSV3030, Rhode \& Schwarz). 

MW signals are generated by a signal generator (SMB100, Rohde \& Schwarz), and the frequency is multiplied by 12 using a frequency multiplier (WR6.5AMC-I, Virginia Diodes Inc.) at room temperature, from which the signal is transmitted to the cryogenic stage via a rectangular waveguide.

\section{Reflectance change\label{sec:reflectance_change}}

A parallel LC circuit weakly coupled to the external environment via a coupling capacitor $C_c$ as shown in Fig.~\ref{fig1}(a) has a reflection coefficient given by~\cite{Oakes2023-ys,Kamigaito2020,Blais2021-zc,Schuster2007-st}
\begin{equation}
    \Gamma_\mathrm{ref}=1 - \frac{2Q_\mathrm{tot}/Q_\mathrm{ext}}{1+j2Q_\mathrm{tot} \left(\frac{f}{f_0} -1  \right)},
    \label{eq:Gamma_quality_factor}
\end{equation}
where the internal quality factor is  $Q_\mathrm{int} = \omega_0 C_\mathrm{t} R$, the external quality factor is $Q_\mathrm{ext} = C_\mathrm{t} / (Z_0 \omega_0 C_c^2)$, the total quality factor is then expressed as 
\(
Q_\mathrm{tot} = \omega_0 C_\mathrm{t} \left( \frac{1}{R} + \frac{(\omega_r C_c Z_0)^2}{Z_0} \right)^{-1}.
\) Here, $R$ represents the circuit loss, which is obtained through fitting. The resonance frequency is $f_0=\omega_0/2\pi =1/2\pi \sqrt{LC_\mathrm{t}}$ with $C_\mathrm{t}=C+C_\mathrm{c}$.

When the total capacitance $C_\mathrm{t}$ changes to $C_\mathrm{t}+\Delta C$, the difference in the reflection coefficient at the probe frequency $f_0$ is given by
\begin{equation}
    \begin{aligned}
        \Gamma_\mathrm{ref}(C_\mathrm{t}+\Delta C) -  \Gamma_\mathrm{ref}(C_\mathrm{t}) 
        & \approx- j\frac{4Q_\mathrm{tot}^2 \left( \frac{f_0'}{f_0} -1   \right) / Q_\mathrm{ext}}{1 + \left(2Q_\mathrm{tot} \left( \frac{f_0'}{f_0} -1  \right)\right)^2} \\
        & \approx -j\frac{2Q_\mathrm{tot}^2}{Q_\mathrm{ext}} \frac{\Delta C }{C_\mathrm{t}},
    \end{aligned}
    \label{eq:GammaDiff}
\end{equation}
where the shifted resonance frequency is given by
$
    f_0'  = \frac{1}{2 \pi  \sqrt{ L (C_\mathrm{t}+\Delta C)}}.
$

\section{Quantum capacitance \label{sec:quantum_C}}

Near zero-detuning \( \epsilon = h(f_\mathrm{Ry} - f_\mathrm{MW}) \sim 0 \), the eigenstates \( \ket{n_z = 1} \) and \( \ket{n_z = 2} \) transform into \( \ket{-} \) and \( \ket{+} \) (see Fig.~\ref{fig1}(c) of the main text). The population of the state \( \ket{-} \) is given by  \( P_-=(1+\chi) /2 \) and that of \( \ket{+} \) \( P_+ = (1-\chi) /2 \), where the difference in population is defined as 
\begin{equation}
\chi = P_- - P_+=\tanh \left( \frac{\Delta E}{2 k_B T} \right)
\end{equation}
for $\epsilon \sim 0$, where $T$ is the electron temperature and $\Delta E =E_+-E_-= \sqrt{\epsilon^2 + (2t_c)^2}$ is the energy difference between \( \ket{-} \) and \( \ket{+} \). \( 2 t_c \) is the Rydberg transition rate. The change in charge induced on the top central electrode by a single electron, relative to \( \ket{n_z = 1} \), is given by
\begin{equation}
    Q_1 = \Delta q \left(|\braket{n_z=2|-} |^2 P_-+ |\braket{n_z=2|+} |^2 P_+  \right),
\end{equation}
where $\Delta q$ is the induced charge difference between $\ket{n_z = 2}$ and $\ket{n_z = 1}$. Knowing that the probability of finding the states $\ket{\pm}$ in $\ket{n_z = 2}$ is given by  $|\braket{n_z=2|\pm} |^2=\frac{1}{2} \left( 1 \mp \frac{\epsilon}{\Delta E}   \right) $, $Q_1$ can be rewritten as
\begin{equation}
    Q_1(\epsilon)= \Delta q \left( \frac{1}{2} +\chi \frac{\epsilon}{2 \Delta E} \right) .
\end{equation}
Thus, the capacitance change by a single electron is given by
\begin{equation}
     C_1 (\epsilon) = \frac{dQ_1}{dV} = \Delta q \frac{dQ_1}{d\epsilon} = C_\mathrm{quantum} + C_\mathrm{tunnel},
\end{equation}
where we used $d\epsilon = \Delta q dV$. The quantum capacitance is defined as 
$C_\mathrm{quantum} = \chi \Delta q^2 \frac{\partial}{\partial \epsilon} \frac{\epsilon}{2 \Delta E}$, and the tunneling capacitance is given by $C_\mathrm{tunnel} = \Delta q^2 \frac{\epsilon}{2 \Delta E} \frac{\partial \chi}{\partial \epsilon}$. The tunneling capacitance becomes non-negligible when population redistribution processes, such as relaxation, occur at a rate comparable to or faster than the probing frequency \cite{Gonzalez-Zalba2016-Ga}. In our case, since the probing frequency 120~MHz is much higher than the relaxation rate 1~MHz, the tunneling capacitance can be ignored. Thus, the capacitance induced by a single electron becomes
\begin{equation}
     C_1 (\epsilon) \approx C_\mathrm{quantum} = \chi \Delta q^2\frac{(2t_c)^2}{2\Delta E ^3} .
\end{equation}
The charge induced by many electrons is given by $Q_N = \int Q_1(\epsilon^0 -\epsilon) n(\epsilon) d\epsilon$ with the electron number distribution $n(\epsilon)$. Therefore, the capacitance change by many electrons is given by 
\begin{equation}
     C_N(\epsilon^0) =  \Delta q \frac{dQ_N}{d\epsilon^0}= \int C_1(\epsilon^0 -\epsilon) n(\epsilon) d\epsilon.
\end{equation}
Now, we introduce MW frequency modulation. Due to this modulation, the detuning becomes time-dependent: $\epsilon^0 (t) = h (f_\mathrm{Ry}^0 - f_\mathrm{MW}^c - f_\mathrm{ma} \cos(2\pi f_\mathrm{mf}t))$. For sufficiently small \( f_\mathrm{ma} \) (i.e., below 470~MHz; see Fig.~\ref{fig4}), as the quantum capacitance is linearly proportional to $f_\mathrm{ma}$ the following approximation holds:
\begin{align}
C_N(t) & = C( h (f_\mathrm{Ry}^0 - f_\mathrm{MW}^c - f_\mathrm{ma} \cos(2\pi f_\mathrm{mf}t)) ) \\
& \approx   C_0 + \delta C \cos(2\pi f_\mathrm{mf}t),
\end{align}
where \( C_0 = C_N( h (f_\mathrm{Ry}^0 - f_\mathrm{MW}^c )) \) and

\begin{equation}
    \delta C = -h f_\mathrm{ma} \frac{dC_N(\epsilon^0)}{d\epsilon} \Big|_{\epsilon^0 = h (f_\mathrm{Ry}^0 - f_\mathrm{MW}^c )}. \label{eq:delta_C}
\end{equation}
Using Eq.~\ref{eq:GammaDiff} and by inserting $\Delta C= C_N(t)$, the reflected signal \( V_\mathrm{o}= \Gamma_\mathrm{ref} (C_\mathrm{t}+C_N(t))V_\mathrm{RF} \cos ( 2\pi f_\mathrm{RF} t) \) can be rewritten as

\begin{equation}
    \begin{aligned}
        V_\mathrm{o} \approx &\ \Gamma_\mathrm{ref} (C_\mathrm{t} )V_\mathrm{RF} \cos ( 2\pi f_\mathrm{RF} t)  \\
        & -j\frac{2Q_\mathrm{tot}^2}{Q_\mathrm{ext}} \frac{C_0 }{C_\mathrm{t}} V_\mathrm{RF}\cos ( 2\pi f_\mathrm{RF} t) \\
        &  -j\frac{2Q_\mathrm{tot}^2}{Q_\mathrm{ext}} \frac{\delta C }{C_\mathrm{t}} V_\mathrm{RF} \cos(2\pi f_\mathrm{mf}t) \cos ( 2\pi f_\mathrm{RF} t).
    \end{aligned}    \label{eq:Vo}
\end{equation}

From Eq.~\ref{eq:delta_C} and Eq.~\ref{eq:Vo}, the signal containing the frequencies \( f_\mathrm{RF} \pm f_\mathrm{mf} \), which corresponds to the sideband peak, is given by

\begin{equation}
    \frac{Q_\mathrm{tot}^2}{Q_\mathrm{ext}} \frac{|\delta C| }{C_\mathrm{t}} V_\mathrm{RF} = \frac{Q_\mathrm{tot}^2}{Q_\mathrm{ext}} \frac{V_\mathrm{RF} }{C_\mathrm{t}}  h f_\mathrm{ma} \left| \frac{dC_N(\epsilon^0)}{d\epsilon} \Big|_{\epsilon^0 = h (f_\mathrm{Ry}^0 - f_\mathrm{MW}^c )} \right|.
    \label{eq:sidepeak}
\end{equation}
Note that the Ryeberg transition can also be detected under continuous-wave (CW) microwave irradiation without FM~\cite{jennings_inprep}. In this case, Eq.~\ref{eq:Vo} reduces to a form without the last term. At \( f = f_0 \), the reflection signal is purely real (first term of Eq.~\ref{eq:Vo}), while the contribution from quantum capacitance appears as a small change in the imaginary part (second term). As a result, amplitude changes are subtle and difficult to resolve using amplitude-based measurements, necessitating phase-sensitive techniques such as lock-in detection. In contrast, FM produces sideband peaks at distinct frequencies, allowing changes in quantum capacitance to be detected via amplitude measurements. Furthermore, this approach enables systematic control of capacitance modulation, making it particularly well suited for evaluating sensitivity.

\section{Simulation of Electric Field Distribution and Rydberg transition frequency\label{sec:Ez_fRy_distribution}}

We computationally modeled two dimensional electrons system using a rectangular grid in cylindrical coordinates $(r, z, \theta)$. Owing to the cylindrical symmetry of the system, the $\theta$-dependence is omitted, and the simulation is carried out in the $r$–$z$ plane. 

The simulation domain spans 2~mm in the vertical direction, corresponding to the space between the bottom-plate and top-plate electrodes, and 7.5~mm in the radial direction, corresponding to the radius of the outer electrode of the Corbino electrodes. The vertical ($z$) direction is discretized into 200 segments and the radial ($r$) direction into 500 segments. This results in a grid resolution of $2~\mathrm{mm}/200$ in $z$ and $7~\mathrm{mm}/500$ in $r$.

The relaxation method was used to calculate the Green's function~\cite{Wilen1988-ui} of an electron at any radial position at $z = 1$~mm (i.e., at grid index $N = 100$). The electric field generated by a surface electron (SE) located at $(r, z) = (r_M, 1~\mathrm{mm})$ is then calculated, where $r_M = 7.5~\mathrm{mm} \times M/500$.

\begin{figure}[ht!]
    \centering
    \includegraphics[width=\linewidth]{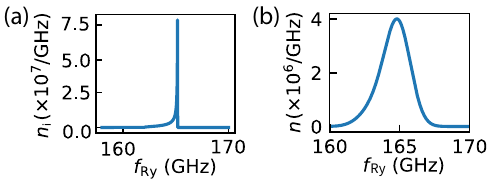}
    \caption{(a) Estimated distribution of the number of electrons $n_i(f_\mathrm{Ry})$ as a function of the Rydberg transition frequency $f_\mathrm{Ry}$, arising from spatial variations of the vertical electric field $E_z$ across the cell. (b) Distribution of the number of electrons $n$ as a function of the Rydberg transition frequency $f_\mathrm{Ry}$.}
    \label{fig:Scewed_f_Ry_distribution}
\end{figure}

The saturated electron density $n_\mathrm{s}$ is determined by adding electrons to the disc until the vertical component of the electric field produced by the SEs exactly cancels the vertical component of the electric field from the electrodes, $E_z$, at a height of $z = 1$~mm. Figure~\ref{fig2}(a) shows the numerically calculated $n_\mathrm{s}$ and $E_z$ as a function of $r$.
The Stark shift induced by $E_z$ is evaluated by numerically solving the Schr\"{o}dinger equation. The number of electrons in each radial disc segment is calculated over the entire grid, leading to an estimate of the broadening $n_\mathrm{i}(f_\mathrm{Ry})$ due to electric field variation across the cell (Fig.~\ref{fig:Scewed_f_Ry_distribution}(a)).
This is then convolved with a Gaussian distribution of 1~GHz width to account for inhomogeneous broadening from other sources (e.g., spatial variations in $E_z$ due to a non-horizontal liquid helium surface or distortions in the top and bottom electrodes), resulting in the total electron distribution $n(f_\mathrm{Ry})$, as shown in Figure~\ref{fig:Scewed_f_Ry_distribution}(b).
Finally, a convolution with $C_1$ is applied to obtain $C_N$ using Eq.~\ref{eq:C_1} and Eq.~\ref{eq:C_N}, where $2t_c/h = 0.83$~MHz is Rydberg transition rate and $T=160$~mK is the temperature, resulting in Fig.~\ref{fig2}(b).

\section{MW power dependence}\label{sec:power_dependence}

\begin{figure}[ht!]
\centering
\includegraphics[width=1\linewidth]{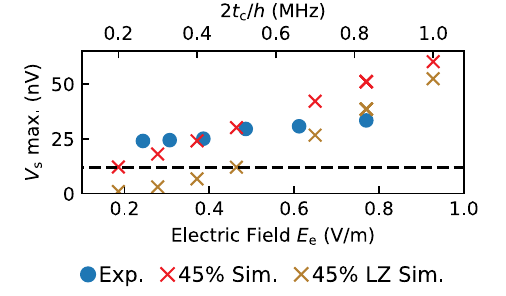}
\caption{Maximum sideband amplitude as a function of the Electric field irradiated to the electrons $E_\mathrm{e}$ (bottom axis) and the corresponding Rydberg transition rate \( 2t_c /h\) (top axis). Blue points (red crosses) represent the experimental (simulation) data. Bronze crosses represent the simulation results including the LZ effects and the dashed black line is the noise level $V_\mathrm{n}=12$~nV. Measurement conditions: \( f_\mathrm{mf} = 1 \)~kHz, \( f_\mathrm{ma} = 768 \)~MHz, and \( V_\mathrm{RF} = 9~\mu\mathrm{V} \).}
\label{fig:power_dependence}
\end{figure}

Figure~\ref{fig:power_dependence} shows the maximum sideband peak amplitude as a function of the corresponding electric field irradiated on the surface electrons (SEs), obtained by varying the microwave power sent from room temperature. The microwave power was controlled using a room-temperature variable attenuator. For all experiments presented in the main text, the attenuation was fixed at $-30~\mathrm{dB}$, corresponding to the highest power shown in Fig.~\ref{fig:power_dependence}. At lower power, nonadiabatic transitions become more prominent, leading to a reduced signal relative to simulations that do not include LZ effects. However, such a reduction was not observed in our experiments. The discrepancy may arise from imperfections in power delivery, such as microwave reflections reducing the attenuator’s effectiveness or the signal approaching the noise floor at high attenuation levels, making accurate power control difficult.

The applied MW intensity cannot be directly measured at low temperature but a calibration was performed with the attenuator at room temperature. The MW intensity was measured as a function of the MW frequency using a WR-6 waveguide detector on the opposite side of the experimental cell, a distance of $r_\mathrm{d}=31.26$~mm from the MW waveguide. Using the area of the WR-6 waveguide, the intensity at the detector for the highest MW power applied in Fig.\ref{fig:power_dependence} is $I_\mathrm{d} = 0.73$~mW/m$^2$. In the far-field regime, this gives an electric field of $E_\mathrm{d}=\sqrt{2 I_\mathrm{d} Z_\mathrm{vac}}$, where $Z_\mathrm{vac}=377~\Omega$ is the impedance of free space . The distance from the waveguide to the electrons at the center of the disc $r_\mathrm{e}=11.26$~mm, and the electric field at the electrons is then $E_\mathrm{e}=E_\mathrm{d}\times \frac{r_\mathrm{d}}{r_\mathrm{e}} = 2$~V/m. 

We can compare this to the $E_\mathrm{e}$ calculated from the Rydberg transition rate~\cite{Platzman1999} as $E_\mathrm{e} =  2t_c/e z_{12}=0.77$~V/m, where $z_{12}=4.6$~nm is the Rydberg transition moment\cite{Platzman1999}. The electric field calculated from the Rydberg transition rate being within an order of magnitude of the calibration measurement is reasonable as the calibration was performed at room temperature in air, the MW waveguide is in a closed cell between two sets of electrodes which have a distance less than the Fraunhofer distance, causing several standing wave resonances, and the electrons may not be at the exactly same solid angle as the detector with regard to the waveguide.

Using higher microwave power than that employed in this work is challenging with the present setup due to heating effects. Moreover, stronger microwave fields can induce excitations to higher Rydberg states~\cite{Kawakami2021}, rendering the two-level approximation invalid.


\section{Comparison with Conventional Detection Techniques and Future improvements\label{sec:appendix_compare}}

Here, we compare the RF reflectometry method devel-
oped in this work with two previously demonstrated tech-
niques: detection as a current in Ref.~\onlinecite{Kawakami2019} and as a voltage in Ref.~\onlinecite{Kawakami2021}. The current measurement described in Ref.~\onlinecite{Kawakami2021} operates at a signal frequency of around 250\,kHz. As the frequency increases, the impedance of the cable’s capacitance decreases, causing the signal to leak to ground and making it difficult to extract. Therefore, the measurement frequency must remain low. However, low-frequency measurements typically suffer from high noise levels, and it is generally difficult to find low-noise cryogenic amplifiers operating in this frequency range. Nevertheless, it remains the simplest and most direct approach. By contrast, the voltage measurement scheme in Ref.~\onlinecite{Kawakami2021} exhibits a broadband response that remains flat between 500\,kHz and 100\,MHz~\cite{Elarabi2021-mb}. This makes the voltage-based approach more suitable for real-time signal detection. In contrast, RF reflectometry is not inherently broadband, but it is better suited for future applications targeting single-electron detection. In the following, we compare the signal amplitude obtained with the voltage-based approach to that of our present RF reflectometry experiments.

For simplicity, we consider the voltage signal induced by a single electron at $\epsilon=0$. Using the RF reflectometry, the quantum capacitance of a single electron is given by Eq.~\ref{eq:C_1}. At $\epsilon=0$ and $k_\mathrm{B} T \gg  t_\mathrm{c}$, this reduces to

\begin{equation}
    C_1 = \Delta q^2 \frac{1}{4 k_\mathrm{B} T}.
\end{equation}

Substituting into Eq.~\ref{eq:Vs}, the voltage signal for the RF reflectometry scheme becomes:
\begin{equation}
    V_\mathrm{s} = \left( G  Q_\mathrm{int}\frac{\Delta q V_\mathrm{RF}}{16k_B T} \right) \frac{ \Delta q}{C_\mathrm{t}} \label{eq:V_s_1}
\end{equation}
at critical coupling. In contrast, in Ref.~\onlinecite{Kawakami2021}, the signal voltage is simply:
\begin{equation}
    V_\mathrm{s} = \frac{ \Delta q}{C_\mathrm{t}}.
\end{equation}
Note that in Ref.~\onlinecite{Kawakami2021}, the impedance matching circuit includes attenuation and amplification stages that compensate each other, resulting in a net gain close to unity. In our present experimental conditions, the prefactor \(
    G  \frac{Q_\mathrm{tot}^2}{Q_\mathrm{ext}} \frac{\Delta q V_\mathrm{RF}}{4k_B T}
\) is approximately 0.01. Therefore, the sensitivity of RF reflectometry is currently lower. 
However, in future implementations aimed at single-electron detection, a different device architecture would be used. Instead of a bulky parallel-plate capacitor, a nanofabricated structure would be employed, allowing the electron to be positioned much closer to a nanofabricated electrode~\cite{Kawakami2023-vf}. As a result, $\Delta q$ can be increased from the present value of $10^{-5}e$ to approximately $10^{-2}e$. Additionally, lowering the temperature from 160\,mK to 10\,mK enhances the prefactor to $\sim 100$, indicating that RF reflectometry will become the superior approach in that regime. Although the device geometry changes, the total capacitance—primarily set by large structures such as bonding pads—is expected to remain comparable. Thus, the LC circuit sensitivity estimated in this work should remain applicable. Note that in conventional semiconductor quantum dot experiments, the typical charge variation is 
\(\Delta q \approx 0.01\text{--}0.89e\)~\cite{Johansson2006-wv,Colless2013-fc,Gonzalez-Zalba2016-Ga,Betz2015-zp,Ibberson2021-uq,Ahmed2018-he} 
depending on the device geometry. The operating temperature is typically around 10~mK, and the relevant capacitance 
\(C_1\) is on the order of a few to hundreds of attofarads~\cite{Colless2013-fc,Chorley2012-aj,Gonzalez-Zalba2015,Ahmed2018-he}. 
The above-discussed future implementations aimed at single-electron detection using electrons on helium are expected 
to approach the lower end of these \(C_1\) values.

The voltage signal in Eq.~\ref{eq:V_s_1} is proportional to both the resonance frequency \( f_0 \) and  \( R \) since \( Q_\mathrm{int} = \omega_0 C_\mathrm{t} R \) and thus by increasing these factors we can also enhance the voltage signal. In this work's setup, an 80\,MHz cutoff was chosen (Fig.~\ref{fig:circuit_details}) to enable Sommer--Tanner measurements~\cite{Sommer1971} between the middle and outer electrodes for confirming electron deposition. Lower cutoff filters are preferable in future implementations to suppress circuit loss.  In addition, replacing the Corbino electrodes—fabricated on an FR4 substrate—with a low-loss, RF-compatible substrate such as Rogers may further reduce circuit loss and enhance \( R \). Increasing the resonance frequency \(f_0\) requires reducing the inductance \(L\), as the total capacitance \(C_\mathrm{t}\)—which would be dominated by the bonding pads—is difficult to reduce further.  However, the achievable \(f_0\) is ultimately limited by the self-resonance of the inductor, which arises from its parasitic capacitance.  Moving to higher frequencies will eventually necessitate a transition from a lumped-element design to a distributed circuit. This shift prohibits multiplexing and eliminates one of the key advantages of the present method—its potential for compact integration~\cite{Jennings2024-sb}.

\bibliography{library}

\end{document}